\documentclass[11pt]{article}
\usepackage[english]{babel}
\usepackage{amsmath,amssymb,mathtools,amsthm}
\usepackage{cite}
\usepackage{xcolor}

\allowdisplaybreaks

\newcommand{\half}{\frac{1}{2}}
\newcommand{\ud}{\mathrm{d}}

\newcommand{\R}{\mathbb{R}}
\newcommand{\N}{\mathbb{N}}
\renewcommand{\vec}{\mathbf}
\renewcommand{\epsilon}{\varepsilon}
\renewcommand{\imath}{\mathrm{i}}

\begin{document}
\title{Noether symmetries and the quantization of a Li\'{e}nard-type nonlinear oscillator}
\author{G. Gubbiotti$^1$ and M.C. Nucci$^{1,2}$}

\date{$^1$ Dipartimento di Matematica e Informatica, Universit\`a di Perugia,
06123 Perugia, Italy\\[0.2cm]
$^2$ INFN Sezione Perugia, Via Alessandro Pascoli, 06123 Perugia, Italy}
%\ead{}
\maketitle
\begin{abstract}
The classical quantization of  a Li\'{e}nard-type nonlinear oscillator is
achieved by a quantization scheme (M.~C. Nucci. {\em Theor. Math. Phys.},
168:997--1004, 2011) that preserves the Noether point symmetries of the
underlying Lagrangian in order to construct the Schr\"odinger equation. This
method straightforwardly yields the correct Schr\"odinger equation in the
momentum space (V.~Chithiika~Ruby, M.~Senthilvelan, and M.~Lakshmanan. {\em J.
Phys. A: Math. Gen.}, 45:382002, 2012), and sheds light into the apparently
remarkable connection with the linear harmonic oscillator.
\end{abstract}
%\pacs{02.30.Hq, 02.20.Sv, 45.20.Jj, 03.65.Ge}

\section{Introduction}

The classical method in the passage from classical to quantum mechanics is
based on the substitution of the classical coordinates and momenta
$(q_{i},p_{i})_{i=1,\ldots,N}$ with the quantum operators:
\begin{equation}
q_{i} \to q_{i}, \quad p_{i} \to -\imath \frac{\partial}{\partial q_{i}}.
\label{correspondence}
\end{equation}
However if the quantization of nonstandard Hamiltonians is pursued then
ambiguity may raise due to ordering non-commuting factors. In such cases, the
normal ordering method -- as described in classical textbooks such
\cite{Bjorken1964,Louisell1990} -- and the Weyl quantization scheme
\cite{Weyl1927} were devised.

Also since quantum mechanics is essentially a \emph{linear} theory then
problems arise when \emph{nonlinear} canonical transformations are involved and
there is no guarantee of consistency
\cite{vanHove1951,Moshinsky78,Anderson1994,Brodlie2004}. For a more recent
perspective on the canonical transformations in quantum mechanics see
\cite{Blaszak2013} and references within.

In \cite{c-iso} it was inferred that Lie symmetries should be preserved if a
consistent quantization is desired. In \cite{Goldstein80} [ex. 18, p. 433] an
alternative Hamiltonian for the simple harmonic oscillator was presented. It is
obtained by applying a nonlinear canonical transformation to the classical
Hamiltonian of the harmonic oscillator. That alternative Hamiltonian was used
in \cite{Nucci2013} to demonstrate what nonsense the usual quantization schemes
produce. In \cite{gallipoli10} a quantization scheme that preserves the Noether
symmetries was proposed and applied  to Goldstein's example in order to derive
the correct Schr\"odinger equation. In \cite{Nucci2012} the same quantization
scheme was applied in order to quantize Calogero's goldfish system.

Let us recall here the quantization scheme that preserves the Noether
symmetries. We consider a linearizable system of second-order ordinary
differential equations, i.e.:
\begin{equation}
\vec{\ddot{x}}(t)= \vec{f}(t,\vec{x},\vec{\dot{x}}), \quad \vec{{x}}\in\R^N
\label{systsec}
\end{equation}
which  possess the maximal number of admissible Lie symmetries, namely
$N^2+4N+3$. In \cite{Gonzalez1983,Gonzalez1988} it was proven that the
maximal-dimension Lie symmetry algebra of a system of $N$ equations of second
order is isomorphic to $sl(N+2,\R)$, and that the corresponding Noether
symmetries generate a $(N^2+3N+6)/2$-dimensional Lie algebra $g^V$ whose
structure (Levi-Mal\'cev decomposition and realization by means of a matrix
algebra) was determined.

The new algorithm that yields the  Schr\"odinger equation can be summarized as
follows:
\begin{description}
\item[Step 1.] Find the linearizing transformation which
do not change the time, as prescribed in non-relativistic quantum mechanics:
\begin{equation}
t \to t, \quad \vec{x} \to \vec{y} = \vec{y}(t,\vec{x}).
\end{equation}
\item[Step 2.]
Find a Lagrangian that admits the maximum number of Noether symmetries, namely
$(N^2+3N+6)/2$.
\item[Step 3.] Apply the linearizing transformation
$\vec{y}=\vec{y}(t,\vec{x})$ to the Schr\"{o}dinger equation  of the
corresponding classical linear problem. This yields the Schr\"{o}dinger
equation corresponding to system \eqref{systsec}.
\item[Step 4.]
Check if the quantization is consistent with the classical properties of the
system, namely find the Lie symmetries of the obtained Schr\"{o}dinger
equation:
\begin{equation}
\Omega = T\partial_{t} + \sum_{i=1}^{n}X_{i} \partial_{x_{i}} + \Psi
\partial_{\psi}.
\end{equation}
and verify that $T\partial_{t} + \sum_{i=1}^{n}X_{i} \partial_{x_{i}}$ generate
the Noether symmetries admitted by the Lagrangian of system \eqref{systsec}.
\end{description}
\vspace{12pt} \noindent \textbf{Remark:} Since the Schr\"{o}dinger equation is
homogeneous and linear, it admits also the  the homogeneity symmetry
$\psi\partial_{\psi}$, and the linearity symmetry $\Omega_{\varPsi}=\varPsi
\partial_\psi$, where $\varPsi$ is any particular solution of the Schr\"{o}dinger
equation itself.\\

In this paper we apply this new quantization algorithm  to a linearizable
Li\'{e}nard equation, i.e.
\begin{equation}
\ddot{x} + k x \dot{x} + \frac{k^2}{9} x^{3} + \omega^{2} x = 0, \label{lien}
\end{equation}
 that has been recently quantized in momentum
space \cite{Senth2012}.

In Section \ref{lienardeq} we recall the properties of the linearizable
Li\'{e}nard equation \eqref{lien}. In Section \ref{invlienardeq} we consider
the classical analogue of the momentum representation of equation \eqref{lien}
as given in \cite{Senth2012} and after showing that is linearizable we quantize
it by preserving the Noether symmetries. In Section \ref{compa} a comparison
between the Noether quantization method and that applied in \cite{Senth2012} is
given. The last Section contains some final remarks.

\section{Classical properties of the Li\'{e}nard equation \eqref{lien}}
\label{lienardeq}

The one-dimensional nonlinear oscillator \eqref{lien} is a special case of the
general Li\'{e}nard equation, i.e.
\begin{equation}
\ddot{x} + f(x) \dot{x} + g(x) = 0, \label{liengen}
\end{equation}
introduced more than 75 years ago \cite{vanderpol1927,Lienard1928} for modeling
electrical circuits. Since then Li\'{e}nard equations have been applied to
many different areas even in biology \cite{Moreira1991}.

The particular Li\'{e}nard equation \eqref{lien}  has been extensively studied
by many authors, the most recent papers being
\cite{Chandrasekar2005,Chandrasekar2006,Chandrasekar2012,Iacono2011}. In
\cite{Mahomed1985,Leach1985,Leach1988} Lie group analysis was applied to
\eqref{liengen} and it was shown that equation \eqref{lien} is the only
linearizable equation among those in \eqref{liengen}. In fact equation
\eqref{lien} was found to admit an eight-dimensional Lie point symmetry algebra
generated by the following operators:
\begin{subequations}
\begin{align}
&\Gamma_{1} = x \partial_{t}-\left(\frac{1}{3} x^3 k + \frac{3 \omega^2}{k} x
\right)\partial_{x},
\\
&\Gamma_{2} = \sin(\omega t) x \partial_{t}-\left(\frac{k}{3} \sin(\omega t)
x^3-\omega \cos(\omega t) x^2 \right)\partial_{x},
\\
&\Gamma_{3} = \cos(\omega t) x \partial_{t}-\left(\frac{k}{3} \cos(\omega t)
x^3 + \omega \sin(\omega t) x^2\right)\partial_{x},
\\
&\begin{aligned}
\Gamma_{4} &= \left(\frac{3 \omega}{k} \cos(2 \omega t) -\sin(2 \omega t) x\right)\partial_{t} \\
&+\left[\left(\frac{k}{3} x^3  - \frac{3 \omega^2}{k} x \right) \sin(2 \omega
t) -2 \omega x^2 \cos(2 \omega t)
 \right]\partial_{x},
\end{aligned}
\\
&\begin{aligned}
\Gamma_{5} &= -\left(\frac{3 \omega}{k} \sin(2 \omega t) +\cos(2 \omega t) x\right)\partial_{t} \\
&-\left[\left(\frac{k}{3} x^3  - \frac{3 \omega^2}{k} x \right) \cos(2 \omega
t) +2 \omega x^2 \cos(2 \omega t)
 \right]\partial_{x},
\end{aligned}
\\
&\Gamma_{6} = \cos(\omega t)\partial_{t} +\left(\omega \sin(\omega t) x-\frac{3
\omega^2}{k} \cos(\omega t) \right)\partial_{x},
\\
&\Gamma_{7} = -\sin(\omega t) \partial_{t}+\left(\omega\cos(\omega t) x+\frac{3 \omega^2}{k}\sin(\omega t)\right)\partial_{x},\\
&\Gamma_{8} = \partial_{t}.
\end{align}
\label{liensimm}
\end{subequations}

Following Lie \cite{Lie12}, the linearizing transformation is given by finding
a  two-dimensional abelian intransitive subalgebra and put it into the
canonical form $\partial_{\tilde x},\;\tilde t\partial_{\tilde x}$. Since a
two-dimensional abelian intransitive subalgebra is that generated by
\begin{equation}
k\Gamma_2-3\omega\Gamma_6,\quad \quad k\Gamma_3-3\omega\Gamma_7
\end{equation}
then the point transformation that takes \eqref{lien} into the one-dimensional
free-particle
\begin{equation}
\frac{{\rm d}^2\tilde x}{{\rm d}\tilde t^2}=0
\end{equation}
is
\begin{equation}
\tilde t=\frac{ k x\cos(\omega t)+3\omega \sin(\omega t)}{k x \sin(\omega
t)-3\omega\cos(\omega t)},\quad\quad \tilde x= -\frac{1}{9\omega^2}\,\frac{x}{
k x\sin(\omega t)-3\omega\cos(\omega t)}. \label{pointtofree}
\end{equation}
Indeed the general solution of \eqref{lien} is known to be
\begin{equation}
x= \frac{9 \omega^3 A\sin(\omega t +\delta)}{k-3\omega^2 k A\cos(\omega
t+\delta)}, \label{liensol}
\end{equation}
with $A$ and $\delta$ arbitrary constants.

Thus equation \eqref{lien} represents a non-linear oscillator -- at least if
$|A| < 1/(3\omega^2)$ -- and should be related to the linear harmonic
oscillator. In \cite{Chandrasekar2005} it was shown that the nonlocal
transformation
\begin{equation}
U = x e^{\frac{k}{3}\int x(\tau) \ud \tau} \label{hononloc}
\end{equation}
takes  equation \eqref{lien} into the linear harmonic oscillator:
\begin{equation}
\ddot{U} + \omega^{2} U =0.\label{LHO}
\end{equation}
Moreover the substitution of the general solution \eqref{liensol} into
\eqref{hononloc} yielded the  following canonical transformation between
\eqref{lien} and \eqref{LHO}:
\begin{equation}
x= \frac{U}{1-\displaystyle \frac{k}{3\omega^2}P},\quad p=P\left(
1-\displaystyle \frac{k}{6\omega^2}P\right). \label{CTSent}
\end{equation}

Here we show that \eqref{hononloc} yields the canonical transformation
\eqref{CTSent} without knowing the general solution of \eqref{lien}.

In \cite{Senth2012} the following  Hamiltonian for equation \eqref{lien} was
presented:
\begin{equation}
H =\frac{9 \omega^4}{2 k^2} \left[2-\frac{2 k}{3 \omega^2} p -2 \left(1-\frac{2
k}{3 \omega^2} p\right)^{\half} + \frac{k^2 x^2}{9 \omega^2}\left(1-\frac{2
k}{3 \omega^2} p\right)\right]. \label{hamsenth}
\end{equation}
with the restriction $-\infty < p\leq 3\omega^2/2k$, and also the corresponding
Lagrangian was derived, i.e.:
\begin{equation}
L = \frac{27 \omega^{6}}{2 k^2}\frac{1}{\displaystyle k \dot{x} + \frac{k^2}{3}
x^2 + 3 \omega^2} + \frac{3 \omega^2}{2 k} \dot{x} - \frac{9 \omega^2}{2 k^2}.
\label{lagrsenth}
\end{equation}

We observe that the last two terms of the Lagrangian \eqref{lagrsenth}
represent the total derivative of the function:
\begin{equation}
G = \frac{3 \omega^2}{2 k} x - \frac{9 \omega^4}{2 k^2} t. \label{G}
\end{equation}
This choice  implies that the momentum is  given by the equation:
\begin{equation}
p \equiv \frac{\partial L}{\partial \dot{x}} = -\frac{27 \omega^{6}}{2
k}\frac{1}{\displaystyle \left(k \dot{x} + \frac{k^2}{3} x^2 + 3
\omega^2\right)^2} + \frac{3 \omega^2}{2 k}, \label{momentumsenth}
\end{equation}
and consequently the velocity $\dot{x}$ is given by:
\begin{equation}
\dot{x} = -\frac{k}{3} x^{2} +  3\omega^2\,
\frac{1-\sqrt{1-\displaystyle\frac{2 k}{3 \omega^2} p}}
{k\sqrt{1-\displaystyle\frac{2 k}{3 \omega^2} p}}.
\end{equation}

\vspace{12pt} \noindent \textbf{Remark:} Although the addition of the total
derivative of the particular function $G$ in \eqref{G} may seem useless, it
actually allows to replace the otherwise ambiguous term $\sqrt{-p}$ with
$\sqrt{1-\displaystyle \frac{2 k}{3 \omega^2} p}$.\\

Apart an unessential multiplicative constant and the addition of the total
derivative of  $G$ in \eqref{G}, the Lagrangian \eqref{lagrsenth} was
determined  in \cite{nuctam_3lag} by means  of the Jacobi Last Multiplier
\cite{Jacobi44a,Jacobi44b,Jacobi45,JacobiVD} as a particular case of the
Lagrangian for the general Li\'{e}nard equation \eqref{liengen}, i.e.:
\begin {equation}
L=\left(\dot x+\frac{g(x)}{\alpha f(x)}\right)^{2-1/\alpha}+\frac{{\rm d}}{{\rm
d}t}G(t,x). \label{Ls}
\end{equation}
when the following relationship holds between $f(x)$ and $g(x)$:
\begin {equation}
\frac{{\rm d}}{{\rm d}x}\left(\frac{g(x)}{f(x)}\right)=\alpha(1-\alpha)f(x),
\label{3}
\end{equation} with $\alpha$ a constant $\neq 1$.

In the case of equation \eqref{lien} it was shown in \cite{nuctam_3lag} that
the relationship \eqref{3} holds, and consequently the following function was
determined:
\begin{equation}
q=\frac{x}{\displaystyle\dot x+\frac{k}{3}x^2+\frac{3\omega^2}{k}}
\label{hocont}
\end{equation}
 such that $q = x e^{\frac{k}{3}\int x(\tau)\ud
\tau}$ and satisfies the linear harmonic oscillator equation:
\begin{equation}
\ddot q +\omega^2 q =0.
\end{equation}

Therefore the nonlocal transformation \eqref{hononloc} is  a contact
transformation and indeed canonical. Substituting $p$ as given in
\eqref{momentumsenth} into \eqref{hocont} yields:
\begin{equation}
x = - \frac{3 \omega^2 q}{\displaystyle k\sqrt{1- \frac{2 k}{3 \omega^2}p}}.
\end{equation}
Let us search for the new momentum $s$ by using a generating function of type
$f_{3} = f_{3}(q,p)$ \cite{Goldstein80}. Then the following equation:
\begin{equation}
\frac{\partial f_{3}}{\partial p} \equiv x= - \frac{3 \omega^2 q}{\displaystyle
k\sqrt{1- \frac{2 k}{3 \omega^2}p}},
\end{equation}
yields the generating function:
\begin{equation}
f_3 = Q(q) + 9 \omega^4 q \sqrt{1 - \frac{2 k}{3 \omega^2} p},
\end{equation}
with $Q(q)$ an arbitrary function of $q$. Then substituting $f_3$ into
\begin{equation}
\frac{\partial f_{3}}{\partial q}=s, \end{equation} yields:
\begin{equation} p=-\frac{k^3}{ 54\omega^6} s^2 + \frac{3 \omega^2}{2 k}
\end{equation}
where we have assumed $Q=0$ for the sake of simplicity. These new canonical
variables $q,s$ transform the Hamiltonian \eqref{hamsenth} into the new
Hamiltonian:
\begin{equation}
K = \half \left(\frac{k s}{9 \omega^2} + \frac{\omega^2}{k}\right)^{2} +\frac{9
\omega^2}{2 k^2} q^2
\end{equation}
which is connected to that of the linear harmonic oscillator by means of the
following \emph{linear} canonical transformation:
\begin{equation}
\tilde{s} = \frac{3 \omega^2}{k} s - \frac{9 \omega^2}{k^2}, \quad \tilde{q} =
\frac{k}{3\omega^2} q,
\end{equation}
and thus the canonical transformation that takes \eqref{lien} into the linear
harmonic oscillator is recovered.

\section{Quantization of \eqref{lien} in momentum space}
\label{invlienardeq}

In \cite{Senth2012} the quantization problem of \eqref{lien} was tackled in the
momentum representation since the Hamiltonian \eqref{hamsenth} is quadratic in
$x$. The von Roos' quantization scheme \cite{vonRoos1983,vonRoos1985} was then
applied.

Instead we begin with the classical equation and then apply the Noether
symmetry quantization.

The canonical transformation \begin{equation} (x,p) \to (X,P) = (p,-x)
\label{caninv}
\end{equation} transforms the Hamiltonian
\eqref{hamsenth} into the ``inverted'' Hamiltonian
\begin{equation}
\tilde{H} =\frac{9 \omega^4}{2 k^2} \left[2-\frac{2 k}{3 \omega^2} X -2
\left(1-\frac{2 k}{3 \omega^2} X\right)^{\half} + \frac{k^2 P^2}{9
\omega^2}\left(1-\frac{2 k}{3 \omega^2} X\right)\right]. \label{hamsenthinv}
\end{equation}
The corresponding Lagrangian is:
\begin{equation}
\tilde{L} = \frac{\dot{X}^2}{2 \omega^2 \displaystyle\left(1-\frac{2 k}{3
\omega^2} X\right)} -\frac{9 \omega^4}{2 k^2} \left(1 - \sqrt{1-\frac{2 k}{3
\omega^2} X}\right)^2. \label{lagrinv}
\end{equation}
and its Lagrangian equation is:
\begin{equation}
\ddot{X}=\frac{3 \omega^{4}}{k} \left(1 - \frac{2 k}{3 \omega^2} X - \sqrt{1 -
\frac{2 k}{3 \omega^2} X}\right) -\frac{k \dot{X}^2}{\displaystyle 3
\omega^2\left(1 - \frac{2 k}{3 \omega^2} X\right)}. \label{eqclasslagr}
\end{equation}
Using the REDUCE programs \cite{Nucci1996} we determine that the Lie symmetry
algebra admitted by equation \eqref{eqclasslagr} is generated by the following
 operators:
\begin{subequations}
\begin{align}
&\Xi_{1} = \partial_{t},
\\
&\Xi_{2} = \cos(2 \omega t) \partial_{t} + \sin(2\omega t)
\frac{3\omega^{3}}{k} \left[1 - \frac{2 k}{3 \omega^{2}} X - \sqrt{1 - \frac{2
k}{3 \omega^{2}} X} \right]\partial_{X}
\\
&\Xi_{3}= \sin(2 \omega t)\partial_{t} - \cos(2\omega t) \frac{3\omega^{3}}{k}
\left[1 - \frac{2 k}{3 \omega^{2}} X - \sqrt{1 - \frac{2 k}{3 \omega^{2}} X}
\right]\partial_{X}
\\
&\Xi_{4} = %\sqrt{3} \omega
\sqrt{1-\frac{2 k}{3 \omega^2} X} \cos(\omega t) \partial_{X}
\\
&\Xi_{5} = %\sqrt{3} \omega
\sqrt{1-\frac{2 k}{3 \omega^2} X} \sin(\omega t) \partial_{X}
\\
&\Xi_{6} =
\begin{aligned}[t]
%\sqrt{\frac{3}{2 k}} \omega
&\cos(\omega t) \left(\sqrt{1 - \frac{2 k}{3 \omega^2} X} - 1\right)\partial_{t}\\
%\sqrt{\frac{3}{2 k}}
&+\frac{3\omega^{3}}{k}\sin(\omega t)\left(\sqrt{1 - \frac{2 k}{3 \omega^2} X}
- 2 \right) \left(1 - \frac{2 k}{3 \omega^2} X\right)\partial_{X},
\end{aligned}
\\
&\Xi_{7} =
\begin{aligned}[t]
&\sin(\omega t) \left(\sqrt{1 - \frac{2 k}{3 \omega^2} X} -
1\right)\partial_{t}
\\
&-\frac{3\omega^{3}}{k}\cos(\omega t)\left(\sqrt{1 - \frac{2 k}{3 \omega^2} X}
- 2 \right) \left(1 - \frac{2 k}{3 \omega^2} X\right)\partial_{X},
\end{aligned}
\\
&\Xi_{8} = \sqrt{1 - \frac{2 k}{3 \omega^2} X} \left(1-\sqrt{1 - \frac{2 k}{3
\omega^2} X}\right)\partial_{X}.
\end{align}
\label{symminveq}
\end{subequations}

Obviously equation \eqref{eqclasslagr} is linearizable and the operators
$\Xi_{i}$ give a representation of the Lie algebra $sl(3,\R)$ \cite{Lie12}.
Therefore the Noether symmetry quantization can be applied step by step.

\paragraph{Step 1.}
Let us find the linearizing transformation. A two-dimensional abelian
intransitive  subalgebra is provided by $\Xi_{4}$ and $\Xi_{5}$ and thus the
linearizing transformation that takes \eqref{eqclasslagr} into the free
particle equation: \begin{equation} \frac{\ud^{2} \xi(\tau)}{\ud \tau^{2}}
=0,\label{free}
\end{equation} is given by
\begin{equation}
\tau = \tan (\omega t),\quad \xi = \omega\sqrt{\frac{6}{k}}\frac{\displaystyle
1 - \sqrt{1 - \frac{2 k}{3 \omega^2} X}}{\cos(\omega t)}.
\end{equation}
Unfortunately this transformation involves changing the time $t$. \\However we
recall that the point transformation between the free particle \eqref{free}
 and the linear harmonic oscillator
  \begin{equation}\frac{{\rm d}^2Z(t)}{{\rm d}t^2}+\omega^{2} Z(t)=0
  \end{equation} is
given by:
\begin{equation}
\tau = \tan (\omega t), \quad \xi = \frac{Z}{\cos(\omega t)}.
\end{equation}
Then it is easy to show that the transformation\footnote{We insert a constant
factor $\omega/\sqrt{k}$ in order to give $\eta$ the same dimension as
$\sqrt{x}$.}:
\begin{equation}
\eta = \omega\sqrt{\frac{6}{k}}\left(1 - \sqrt{1 - \frac{2 k}{3 \omega^2}
X}\right), \label{linearizingnotime}
\end{equation}
takes equation \eqref{eqclasslagr} into the linear harmonic oscillator:
\begin{equation}
\ddot{\eta} + \omega^{2} \eta =0. \label{invlinearized}
\end{equation}
 Thus the linearization transformation
\eqref{linearizingnotime} yields the general solution of equation
\eqref{eqclasslagr}, i.e.:
\begin{equation}
X(t) = \frac{3 \omega^2}{2k} \left[1-\left(1-
\frac{1}{\omega}\sqrt{\frac{k}{6}}A\sin(\omega t+\delta)\right)^{2}\right],
\end{equation}
where $A$ and $\delta$ are two arbitrary constants.

\vspace{12pt} \noindent \textbf{Remark:} Equation \eqref{eqclasslagr} and
Li\'{e}nard equation \eqref{lien} are
 examples of nonlinear oscillators whose amplitude do not depend on the
frequency, unlike other famous nonlinear oscillators, e.g. the
Mathews-Lakshmanan oscillator \cite{Mathews1974,Lakshmanan2003}.

\paragraph{Step 2.}
The Lagrangian \eqref{lagrinv} admits five Noether point symmetries, namely
$\Xi_{i}$ with $i=1,\ldots,5$ in \eqref{symminveq}.

\paragraph{Step 3.}
Let us consider the Schr\"{o}dinger equation for the linear  harmonic
oscillator:
\begin{equation}
2 \imath \widetilde\Phi_{t} + \widetilde\Phi_{\eta\eta} - \omega^2
\eta^2\widetilde\Phi =0, \label{schrsymm}
\end{equation}
with $\widetilde\Phi=\widetilde\Phi(t,\eta)$. Then applying the transformation
\eqref{linearizingnotime} we obtain the following Schr\"{o}dinger equation :
\begin{equation}
2 \imath \widetilde\Phi_{t} + \frac{3 \omega^2}{2 k}\left(1-\frac{2
k}{3\omega^2}X\right)\widetilde\Phi_{XX} -\frac{1}{2} \widetilde\Phi_{X}
-\frac{6 \omega^4}{k} \left(1-\sqrt{1-\frac{2k}{3\omega^2}
X}\right)^{2}\widetilde\Phi, \label{schrcompl}
\end{equation}
with $\widetilde\Phi=\widetilde\Phi(t,X)$. The following transformation
eliminates the first derivative of $\Phi$ with respect to $X$ in
\eqref{schrcompl}:
\begin{equation}
{\Phi}(t,X) =
\frac{\widetilde\Phi(t,X)}{\displaystyle\left[3\omega^2\left(1-\frac{2k}{3\omega^2}X\right)\right]^\frac{1}{4}}
\end{equation}
and hence the final form of the Schr\"{o}dinger equation is:
\begin{equation}
\begin{aligned}
2 \imath \Phi_{t} &+ 9
\left(1-\frac{2 k}{3\omega^2}X\right)\Phi_{XX}\\
&+ \left[ \frac{3k^2}{4\omega^4\left(1-\frac{2 k}{3\omega^2}X\right)}
-\frac{2\omega^6}{k^2}\left(1-\frac{k}{3\omega^2}X\right) \sqrt{1-\frac{2
k}{3\omega^2}X} \right]\Phi.
\end{aligned}
\label{schrfinal}
\end{equation}

\paragraph{Step 4.}
We now check the classical consistency of the Schr\"{o}dinger equation
\eqref{schrfinal}. Using the REDUCE programs \cite{Nucci1996} we find that its
Lie point symmetries are generated by the following operators:
\begin{subequations}
\begin{align}
\Omega_{1} &=
\Xi_{1},\\
\Omega_{2} &= \Xi_{2} + \left[\frac{\omega}{2}\sin(2 \omega t)
-\imath\omega^2\cos(2\omega t) \left(1-\sqrt{1-\frac{2 k}{3 \omega^2}
X}\right)^2 \right]
\frac{\Phi}{\displaystyle\sqrt{1-\frac{2 k}{3 \omega^2} X}}\partial_{\Phi},\\
\Omega_{3} &= \Xi_{3} - \left[\frac{\omega}{2}\cos(2 \omega t)
+\imath\omega^2\sin(2\omega t) \left(1-\sqrt{1-\frac{2 k}{3 \omega^2}
X}\right)^2 \right]
\frac{\Phi}{\displaystyle\sqrt{1-\frac{2 k}{3 \omega^2} X}}\partial_{\Phi},\\
\Omega_{4} &=
\Xi_{4}%\sqrt{3}\omega
- \left[ \frac{k\cos(\omega t)}{6 \omega^2 \displaystyle\sqrt{1-\frac{2 k}{3
\omega^2} X}} +2 \imath \omega \left( 1 - \sqrt{1-\frac{2 k}{3 \omega^2} X}
\right)\sin(\omega t) \right] \Phi\partial_{\Phi}
%+ \frac{\imath k}{3\omega} \sin(\omega t)
%\left(\sqrt{1-\frac{2 k}{3 \omega^2} X}-1\right)
%\Phi\partial_{\Phi},\\
\\
\Omega_{5} &= \Xi_{5} - \left[ \frac{k\sin(\omega t)}{6 \omega^2
\displaystyle\sqrt{1-\frac{2 k}{3 \omega^2} X}} -2 \imath \omega \left( 1 -
\sqrt{1-\frac{2 k}{3 \omega^2} X} \right)\cos(\omega t) \right]
\Phi\partial_{\Phi}
% - \frac{\imath k}{3\omega} \cos(\omega t)
%\left(\sqrt{1-\frac{2 k}{3 \omega^2} X}-1\right)
%\Phi\partial_{\Phi},
\\
\Omega_{6} &=
\Phi \partial_{\Phi},\\
\Omega_{\chi} &= \chi \partial_{\Phi},
\end{align}
\label{schrsymmetries}
\end{subequations}
where $\chi$ is any solution of \eqref{schrfinal}.\\

It was shown in \cite{Senth2012} that the spectrum of the Li\'{e}nard equation
in momentum space consists of two parts, one positive and one negative.

The positive part is:
\[
E_{n} = \omega\left(n +\half\right), \quad n\in\N,
\]
which is the spectrum of the quantum harmonic oscillator. The eigenfunctions of
this part of the spectrum satisfy the boundary conditions:
\begin{equation}
\begin{aligned}
\lim_{X\to -\infty}\Phi(t,X)=0, \quad& \text{for every $t\in\R_{+}$},\\
\Phi\left(t,\frac{3\omega^2}{2k}\right) = 0,\quad &\text{for every
$t\in\R_{+}$}.
\end{aligned}\label{posBC}
\end{equation}

The negative part is:
\[
E_{n^-} = -\omega\left(n +\half\right), \quad n\in\N,
\]
which is again the spectrum of the quantum harmonic oscillator, but with a
negative sign. The eigenfunctions of this part of the spectrum  satisfy the
boundary conditions:
\begin{equation}
\begin{aligned}
\lim_{X\to +\infty}\Phi(t,X)=0, \quad& \text{for every $t\in\R_{+}$},\\
\Phi\left(t,\frac{3\omega^2}{2k}\right) = 0,\quad &\text{for every
$t\in\R_{+}$}.
\end{aligned}\label{negBC}
\end{equation}

In \cite{Leach05,KostisLeach05,jlmschqm,Nucci2010} and more recently in
\cite{Nucci2013} it was shown how to find the eigenfunctions and the
eigenvalues of the Schr\"{o}dinger equation by means of its admitted Lie
symmetries.

We apply this method to the Schr\"{o}dinger equation \eqref{schrfinal} and find
the same results as in \cite{Senth2012}.

Let us rewrite the Lie point symmetries \eqref{schrsymmetries} of equation
\eqref{schrfinal} in complex form, i.e.:
\begin{subequations}
\begin{align}
&\Sigma_{1} = \imath \partial_{t},\\
&\Sigma_{2\pm}
\begin{aligned}[t]
&= e^{\pm 2  \imath  \omega t}\partial_t \mp \imath e^{\pm 2  \imath  \omega t}
\frac{3\omega^3}{k} \left(1-\sqrt{1-\frac{2 k}{3\omega^2}
X}\right)\sqrt{1-\frac{2 k}{3\omega^2} X}
\partial_{X}\\
&+ \frac{\imath}{2} e^{\pm 2  \imath  \omega t}
\left[\frac{24 \omega^3-24 \omega^3 \sqrt{1-\frac{2 k }{3 \omega^2}X}-k+8 k X \omega \sqrt{1-\frac{2 k}{3 \omega^2}X}-16 k \omega X }%
{\displaystyle\omega k\sqrt{1-\frac{2 k }{3 \omega^2}X}}\right]
\Phi\partial_{\Phi}
\end{aligned}
\\
&\Sigma_{3\pm}
\begin{aligned}[t]
&=
e^{\pm \imath \omega}%\sqrt{3}\omega
\sqrt{1-\frac{2 k}{3\omega^2} X}\partial_{X}
\\
&\mp e^{\pm \imath \omega^3} 2 \omega\left[ \left(1- \sqrt{1-\frac{2
k}{3\omega^2} X}\right) +\frac{k}{12 \sqrt{3}\omega^4}
\frac{1}{\displaystyle\sqrt{1-\frac{2 k}{3\omega^2} X}}
\right]\Phi\partial_{\Phi}%\half
\end{aligned}
\\
&\Sigma_{4} = \Phi \partial_{\Phi}
\\
&\Sigma_{\chi} = \chi \partial_{\Phi}
\end{align}
\end{subequations}

 The operators $\Sigma_{3\pm}$  act as creation and annihilation
operators, namely in the case of  the boundary conditions \eqref{posBC}
$\Sigma_{3+}$ is the annihilation operator and $\Sigma_{3-}$ is the creation
operator, and viceversa in the case of the boundary conditions \eqref{negBC}.

Let us now consider the case of the boundary conditions \eqref{posBC}.
 The invariant surface of the operator $\Sigma_{3+}$ is given by
\begin{equation}
F(t,X,\Phi) = f\left(t, \Phi
\frac{e^{-\frac{2\omega}{k}\left(3\omega^2\sqrt{1-\frac{2 k}{3\omega^2} X}+k X\right)}}%
{\left(1-\frac{2k}{3\omega^2}X\right)^{\frac{1}{4}}}\right) = 0,
\end{equation}
which if solved with respect to $\Phi$ yields
\begin{equation}
\Phi(t,X) = T(t) \left(1-\frac{2k}{3\omega^2}X\right)^{\frac{1}{4}}
e^{\frac{2\omega}{k}\left(3\omega^2\sqrt{1-\frac{2 k}{3\omega^2} X}+k
X\right)},
\end{equation}
with $T(t)$ arbitrary function of $t$. Substituting this solution into the
Schr\"{o}dinger equation \eqref{schrfinal} yields $T(t) = e^{-\half \imath t}$
and thus the ground state solution is:
\begin{equation}
\Phi_{0}(t,X) = \left(1-\frac{2k}{3\omega^2}X\right)^{\frac{1}{4}} e^{-\half
\imath t+\frac{2\omega}{k}\left(3\omega^2\sqrt{1-\frac{2 k}{3\omega^2} X} +k
X\right)}. \label{phi0}
\end{equation}
This solution $\Phi_{0}(t,X)$ satisfies the boundary conditions \eqref{posBC}
and is indeed the ground state\footnote{Apart from an unessential normalization
constant.} since
\begin{equation}
\left[\Sigma_{3+},\Sigma_{\Phi_{0}}\right] = 0.
\end{equation}
Thus there are no states under $\Phi_{0}(t,X)$.

The operator $\Sigma_{1}$ acts like an eigenvalue operator since:
\begin{equation}
\Sigma_{1} \Phi_{0} = \frac{\omega}{2} \Phi_{0},
\end{equation}
which yields the ground energy $E_{0} = \omega/2$, just like the usual quantum
harmonic oscillator.

We use the creation operator $\Sigma_{3-}$ and $\Sigma_{\Phi_{0}}$ in order to
construct the higher states. Since the commutator:
\begin{equation}
\left[\Sigma_{3-},\Sigma_{\Phi_{0}}\right] = 4\omega^2 e^{-\frac{3}{2} \imath
\omega t+\frac{2\omega}{k}\left(3\omega^2\sqrt{1-\frac{2 k}{3\omega^2} X}+k
X\right)} \left(1-\frac{2k}{3\omega^2}X\right)^{\frac{1}{4}}
\left(\sqrt{1-\frac{2 k}{3\omega^2} X}-1\right)\partial_{\Phi},
\end{equation}
then:
\begin{equation}
\Phi_{1}(t,X)= 4\omega^2 e^{-\frac{3}{2} \imath \omega
t+\frac{2\omega}{k}\left(3\omega^2\sqrt{1-\frac{2 k}{3\omega^2} X}+k X\right)}
\left(1-\frac{2k}{3\omega^2}X\right)^{\frac{1}{4}} \left(\sqrt{1-\frac{2
k}{3\omega^2} X}-1\right)
\end{equation}
is another solution of \eqref{schrfinal} that satisfies the boundary conditions
\eqref{posBC} and has a greater energy eigenvalue $E_{1}=\frac{3\omega}{2}$
given by:
\begin{equation}
\Sigma_{1} \Phi_{1} = \frac{3\omega}{2} \Phi_{1}.
\end{equation}

If we evaluate the commutator between $\Sigma_{3+}$ and $\Sigma_{\Phi_{1}}$
then we obtain a multiple of $\Phi_{0}$, i.e.:
\begin{equation}
\left[\Sigma_{3+},\Sigma_{\Phi_1}\right] = -4k\omega \Phi_{0}\partial_{\Phi},
\end{equation}
and thus we have constructed the first excited state.

Iterating the process yields all the eigenfunctions, i.e.:
\begin{equation}
\left[\Sigma_{3-},\Sigma_{\Phi_{n-1}}\right] = \Phi_{n} \partial_{\Phi} =
\Sigma_{\Phi_{n}}. \label{phin}
\end{equation}

Since we have proven that $\Sigma_{3+}$ acts as the annihilation operator for
the first excited state, then we can easily show by means of Jacobi
identity\footnote{And by means of
$$\left[\Sigma_{3+},\Sigma_{3-}\right]=4k\omega\Sigma_4$$.}  that
this holds true for every $n\in \N$, i.e.:
\begin{equation}
\begin{aligned}
\left[\Sigma_{3+},\Sigma_{\Phi_{n}}\right] &=
\left[\Sigma_{3+},\left[\Sigma_{3-},\Sigma_{\Phi_{n-1}}\right]\right]
\\
&= \left[\left[\Sigma_{\Phi_{n-1}},\Sigma_{3+}\right],\Sigma_{3-}\right] +
\left[\left[\Sigma_{3+},\Sigma_{3-}\right],\Sigma_{\Phi_{n-1}}\right]
\\
&= -\kappa\left[\Sigma_{\Phi_{n-2}},\Sigma_{3-}\right] =
\kappa\Sigma_{\Phi_{n-1}}
\end{aligned}
\end{equation}
where $\kappa$ is a constant.

The generic eigenvalue and eigenfunction can be derived in the following
manner. We evaluate the commutator between $\Sigma_{3-}$ and $\Sigma_{\chi}$,
where $\chi$ is a generic solution of \eqref{schrfinal}:
\begin{equation}
\begin{aligned}
 \left[\Sigma_{3-},\Sigma_{\chi}\right] = \frac{e^{-\imath \omega t}}{2\omega
\sqrt{3}\sqrt{1-\frac{2 k}{3\omega^2} X}} &\left[ \left(12 \omega^3 +k - 8 k
\omega X - 12 \omega^3   \sqrt{1-\frac{2 k}{3 \omega^2} X}\right) \chi \right.
\\ &+(6 \omega^2- 4 k X) \chi_{X} \Biggr]
\partial_{\Phi}.
\end{aligned}
\end{equation}

We define the operator $\hat{O}$:
\begin{equation}
\hat{O} =\frac{1}{2\omega \sqrt{3}\sqrt{1-\frac{2 k}{3\omega^2} X}}\left[ 12
\omega^3 + k - 8 k  \omega X - 12 \omega^3   \sqrt{1-\frac{2 k}{3 \omega^2} X}
+(6 \omega^2- 4 k X) \partial_{X} \right]
\end{equation}
and then beginning with the ground state $\Phi_{0}$ generate the $n$th
eigenfunction by using the iteration procedure \eqref{phin}, i.e.:
\begin{equation}
\begin{aligned}
\Phi_{1} &=e^{-\imath \omega t}\hat{O} \Phi_{0},
\\
\Phi_{2} &= e^{-\imath \omega t}\hat{O} \Phi_{1} = e^{-2\imath \omega
t}\hat{O}^{2}\Phi_{0},
\\
&\vdots
\\
\Phi_{n} &=e^{-\imath \omega t}\hat{O} \Phi_{n-1} = e^{-\imath n \omega
t}\hat{O}^{n}\Phi_{0}.
\end{aligned}
\end{equation}

Since the operator $\hat{O}$ acts  on  $X$ only, then
\begin{equation}
\Phi_{n} = e^{-\imath \left(n+\half\right)\omega t} \hat{O}^{n} \left(
\left(1-\frac{2k}{3\omega^2}X\right)^{\frac{1}{4}}
e^{\frac{2\omega}{k}\left(3\omega^2\sqrt{1-\frac{2 k}{3\omega^2} X}+k X\right)}
\right).
\end{equation}
Consequently,  applying the eigenvalue operator $\Sigma_{1}$ yields the
positive part of the spectrum, i.e.:
\begin{equation}
\Sigma_{1} \Phi_{n} = \omega\left(n+\half\right) \Phi_{n}.
\end{equation}

The negative part of the spectrum can be determined in the same way. We
determine the invariant surface of $\Sigma_{3-}$, i.e.:
\begin{equation}
G(t,X,\Phi)=g\left(t,\Phi
\frac{e^{\frac{2\omega}{k}\left(3\omega^2\imath\sqrt{\frac{2 k}{3\omega^2} X-1}+k X\right)}}%
{\left(\frac{2k}{3\omega^2}X-1\right)^{\frac{1}{4}}}\right)
\end{equation}
that solved with respect to $\Phi$ yields:
\begin{equation}
\Phi_{0^{-}}(t,X) = \tilde{T}(t)
\left(\frac{2k}{3\omega^2}X-1\right)^{\frac{1}{4}}
e^{-\frac{2\omega}{k}\left(3\omega^2\imath\sqrt{\frac{2 k}{3\omega^2} X-1}+k
X\right)}.
\end{equation}
Substituting $\Phi_{0^{-}}(t,X)$ into \eqref{schrfinal} yields $\tilde{T}(t)=
e^{\frac{\omega}{2} \imath t}$, i.e. the solution:
\begin{equation}
\Phi_{0^{-}}(t,X) = \left(\frac{2k}{3\omega^2}X-1\right)^{\frac{1}{4}}
e^{\frac{\omega}{2} \imath t-\frac{2\omega}{k}%
\left(3\omega^2\imath\sqrt{\frac{2 k}{3\omega^2} X-1}+k X\right)}.
\label{phi0m}
\end{equation}

The function $\Phi_{0^{-}}(t,X)$ \eqref{phi0m} satisfies  the boundary
conditions \eqref{negBC}.

$\Phi_{0^{-}}(t,X)$ is equivalent to the ground state since
\begin{equation}
\left[\Sigma_{3-},\Sigma_{\Phi_{0^{-}}}\right] = 0.
\end{equation}
and thus there are no states above $\Phi_{0^{-}}$.

The operator $\Sigma_{1}$ acts like an eigenvalue operator since:
\begin{equation}
\Sigma_{1} \Phi_{0^{-}} = -\frac{\omega}{2}\Phi_{0^{-}},
\end{equation}
which yields the ground energy $E_{0^{-}} = -\omega/2$, and consequently
$\Phi_{0^{-}}(t,X)$ has a negative eigenvalue.

Since $\left[\Sigma_{3+},\Sigma_{\Phi_{0^{-}}}\right]=\Sigma_{\Phi_{1^{-}}}$,
 we explicitly determine the first negative excited state
$\Phi_{1^{-}}(t,X)$,  i.e.:
\begin{equation}
\Phi_{1^{-}} = 4\sqrt{3}\omega^{\frac{3}{2}} e^{\frac{3}{2} \imath \omega
t-\frac{2\omega}{k}\left(3\omega^2\sqrt{1-\frac{2 k}{3\omega^2} X}+k X\right)}
\left(\frac{2k}{3\omega^2}X-1\right)^{\frac{1}{4}} \left(\imath\sqrt{\frac{2
k}{3\omega^2} X-1}-1\right).
\end{equation}
Applying $\Sigma_{1}$ to $\Phi_{1^{-}}(t,X)$ we get the corresponding
eigenvalue:
\begin{equation}
\Sigma_{1}\Phi_{1^{-}} = -\frac{3\omega}{2}\Phi_{1^{-}},
\end{equation}
and by applying the commutator with $\Sigma_{3-}$ we indeed obtain:
\begin{equation}
\left[\Sigma_{3-},\Sigma_{\Phi_{1^{-}}}\right]=4\omega k\Sigma_{\Phi_{0^{-}}}.
\end{equation}
Finally in analogy with the case of  the positive part of the spectrum, we have
the following recursive formula yielding all the eigenvalues and
eigenfunctions:
\begin{align}
\left[\Sigma_{3+},\Sigma_{\Phi_{n-1^{-}}}\right] &= \Phi_{n^{-}}\partial_{\Phi}
=\Sigma_{\Phi_{n^{-}}},\\
\left[\Sigma_{3-},\Sigma_{\Phi_{n^{-}}}\right] &= \Phi_{n-1^{-}}\partial_{\Phi}
=\Sigma_{\Phi_{n-1^{-}}},\\
\Sigma_{1} \Phi_{n^{-}} &= -\omega\left(n +\half\right) \Phi_{n^{-}}.
\end{align}

\section{Comparison between the two quantization methods for \eqref{eqclasslagr}}
\label{compa}

We compare the Noether symmetry quantization method applied to equation
\eqref{eqclasslagr}, as shown in the previous Section, with that used in
\cite{Senth2012}. Since the Hamiltonian \eqref{hamsenth} is a nonstandard one,
the classical quantization rule \eqref{correspondence} cannot be used.
Therefore in \cite{Senth2012} a simple modification of the quantization scheme
proposed by von Roos \cite{vonRoos1983,vonRoos1985} for position-dependent
masses was applied. Since the Hamiltonian \eqref{hamsenth} assumes the form
\begin{equation}
H=\frac{x^2}{2m(p)} + U(p),
\end{equation}
the authors of \cite{Senth2012} introduced the following momentum-dependent
mass:
\begin{equation}
m(p) = \frac{1}{\omega^2 \left(1-\frac{2 k}{3 \omega^2} p\right)},
\end{equation}
and also the following momentum-dependent potential:
\begin{equation}
U(p) = \frac{9 \omega^4}{2 k^2} \left(\sqrt{1-\frac{2 k}{3 \omega^2} p}
-1\right)^{2}.
\end{equation}
Then the following Schr\"{o}dinger equation was obtained\footnote{In
\cite{Senth2012} time-independent Schr\"{o}dinger equations were derived.}:
\begin{equation}
\begin{aligned}
2 \imath\Psi_{t}  +& \omega^{2}\left(1-\frac{2 k}{3 \omega^2} p\right)
\Psi_{pp} - \frac{2 k}{3}\Psi_{p} \\ &+\left[\frac{4 k^{2}\alpha(\alpha + \beta
+1)}{9 \omega^{2}\left(1-\displaystyle\frac{2 k}{3 \omega^2} p\right)} -
\frac{9 \omega^{4}}{k^{2}}\left(\sqrt{1-\frac{2 k}{3 \omega^2}
p}-1\right)^{2}\right]\Psi
 = 0.
\end{aligned}
\label{schrsenth1}
\end{equation}
where the constants $\alpha$ and $\beta$ are related with the other constant
$\gamma$  by means of the condition
 \begin{equation}
  \alpha + \beta + \gamma=-1\label{vRc}
\end{equation}
 as prescribed by the von Roos' method. Moreover
the following further restriction was imposed:
\begin{equation}
4\alpha(\alpha + \beta +1)=-\frac{1}{4} \label{coeffcond}
\end{equation}
in order to find the solution of equation \eqref{schrsenth1} by means of
Hermite differential equation since the spectrum is that described in the
previous Section.

Unfortunately the eigenfunctions are singular on the right boundary $p= 3
\omega^2/2 k$ and consequently another Schr\"{o}dinger equation was proposed by
considering the following modified Hamiltonian :
\begin{equation}
\bar H=m^{d} H m^{-d},
\end{equation}
and applying to it the von Roos' procedure. Thus it was found that $d$ must be
equal to $-1/2$, and the following Schr\"{o}dinger equation was obtained:
\begin{equation}
\begin{aligned}
2 \imath\Theta_{t}  +& \omega^{2}\left(1-\frac{2 k}{3 \omega^2} p\right)
\Theta_{pp} \\ &+\left[\frac{k^{2}}{12 \omega^{2}\left(1-\displaystyle\frac{2
k}{3 \omega^2} p\right)} - \frac{9 \omega^{4}}{k^{2}}\left(\sqrt{1-\frac{2 k}{3
\omega^2} p}-1\right)^{2}\right]\Theta
 = 0,
\end{aligned}
\label{schrsenth2}
\end{equation}
which has bounded eigenfunctions and the same spectrum of equation \eqref{schrsenth1}.\\

We notice that equations \eqref{schrsenth1} and \eqref{schrsenth2} are actually
related by the point transformation:
\begin{equation}
\Psi(p,t) = \frac{\Theta(p,t)}{\displaystyle\sqrt{3\omega^{2}\left(1 - \frac{2
k}{3 \omega^2} p\right)}}, \label{schrsenthreleq}
\end{equation}
that eliminates the first-derivative in equation \eqref{schrsenth1}. Therefore
the two equations admit the same finite Lie point symmetry algebra.
\label{schrsenthrel}

Moreover, the condition \eqref{coeffcond} on the constants $\alpha$ and $\beta$
is actually equivalent to require that the finite Lie point symmetry algebra of
equation \eqref{schrsenth1} -- and consequently equation \eqref{schrsenth2} --
be six dimensional. Indeed the full Lie symmetry algebra is generated by the
following operators:\begin{subequations}
\begin{align}
&\Gamma_{1} = \Xi_{1}{\Big |}_{X=p},
\\
&\Gamma_{2} =
\begin{aligned}[t]
\Xi_{2}{\Big |}_{X=p}+  &\left[ \frac{\omega}{2}\sin(2 \omega t) \Psi \frac{2
\sqrt{1-\frac{2 k}{3 \omega^2} p} - 1}{\sqrt{1-\frac{2 k}{3 \omega^2} p}}\right.\\
&  \left.- \frac{9\imath\omega^4}{k^2} \cos(2 \omega t) \Psi
 \left(\sqrt{1-\frac{2 k}{3 \omega^2} p}-1\right)^2
\right]\partial_{\Psi},
\end{aligned}
\\
&\Gamma_{3}=
\begin{aligned}[t]
\Xi_{3}{\Big |}_{X=p} - &\left[ \frac{\omega}{2}\cos(2 \omega t) \Psi \frac{2
\sqrt{1-\frac{2 k}{3 \omega^2} p} - 1}{\sqrt{1-\frac{2 k}{3 \omega^2} p}}\right.\\
&  \left. + \frac{9\imath\omega^4}{k^2} \sin(2 \omega t) \Psi
\left(\sqrt{1-\frac{2 k}{3 \omega^2} p}-1\right)^2 \right]\partial_{\Psi}
\end{aligned}
\\
&\Gamma_{4} =\begin{aligned}[t] \Xi_{4}{\Big |}_{X=p} + &\left[ \frac{k}{6
\omega^2}
\cos(\omega t) \frac{\Psi}{ \sqrt{1-\frac{2 k}{3 \omega^2} p}} \right.\\
&  \left. -\frac{3\imath\omega}{k} \sin(\omega t) \Psi \left(1-\sqrt{1-\frac{2
k}{3 \omega^2} p}\right) \right]\partial_{\Psi},
\end{aligned}
\\
&\Gamma_{5} =\begin{aligned}[t] \Xi_{5}{\Big |}_{X=p}+ &\left[ \frac{k}{6
\omega^2}
\sin(\omega t) \frac{\Psi}{ \sqrt{1-\frac{2 k}{3 \omega^2} p}}\right.\\
&  \left. +\frac{3\imath\omega}{k} \cos(\omega t) \Psi \left(1-\sqrt{1-\frac{2
k}{3 \omega^2} p}\right) \right]\partial_{\Psi},
\end{aligned}
\\
&\Gamma_{6} = \Psi \partial_{\Psi},
\\
&\Gamma_{\psi} = \varPsi(p,t) \partial_{\Psi}
\end{align}
\label{schrsenthsymm}
\end{subequations}
where $\varPsi(p,t)$ satisfies equation \eqref{schrsenth1}, and $\Xi_{j}{\Big
|}_{X=p} (j=1,5)\,$ are the generators of the Noether symmetries
\eqref{symminveq} of equation \eqref{eqclasslagr} with $X$ replaced by $p$.

 If $\alpha$ and $\beta$
do not satisfy condition \eqref{coeffcond} then equation \eqref{schrsenth1}
admits  only  $\Gamma_{1}$, $\Gamma_{6}$ and $\Gamma_{\psi}$ as Lie symmetries.

\section{Final remarks}
We have recalled an algorithm for quantization  that requires the preservation
of Noether symmetries in the passage from classical to quantum mechanics
\cite{gallipoli10,Nucci2012}. We have emphasized that it can be applied if the
classical system of $N$ second-order  equations admits the maximum number of
Lie symmetries, namely $N^2+4N+3$, and the linearizing transformation is
time-independent.

Thus the derived Schr\"{o}dinger equation is such that the
independent-variables part of its admitted Lie symmetries corresponds to the
Noether symmetries of the classical equations.

We have  applied this quantization method to the linearizable Li\'enard
equation \eqref{lien} and compared our results with that found in
\cite{Senth2012}.

We have shown that even in quantum mechanics whenever differential equations
are involved, Lie and Noether symmetries have a fundamental role.

Indeed Noether symmetries give raise to the correct Schr\"odinger equation and
its Lie symmetries can be algorithmically used to find the eigenvalues and
eigenfunctions.

\section*{Acknowledgement}
MCN acknowledges the support of the Italian Ministry of University and
Scientific Research through PRIN 2010-2011, Prot. 2010JJ4KPA\_004.

\end{document}